\documentclass[12pt]{article}
\usepackage{a4wide,epsfig}

\newlength{\captionwidth}
\begin{document}    

\title{\vskip-3cm{\baselineskip14pt
\centerline{\normalsize\hfill Freiburg-THEP 00/16}
\centerline{\normalsize\hfill TTP 00-26}
\centerline{\normalsize\hfill hep-ph/0011373}
\centerline{\normalsize\hfill November 2000}
}
\vskip.4cm
Three-Loop Leading Top Mass Contributions \\ to the $\rho$ Parameter
\vskip.4cm
}
\author{
J.J. van der Bij${}^{a}$,
{K.G. Chetyrkin}\thanks{Permanent address:
Institute for Nuclear Research, Russian Academy of Sciences,
60th October Anniversary Prospect 7a, Moscow 117312, Russia.}
${}^{,a,b}$,
{M. Faisst}${}^{b}$,
G. Jikia${}^{a}$,
\\
and
{T. Seidensticker}${}^{b}$
\\[3em]
${}^a${\it Fakult{\"a}t f{\"u}r Physik,}
\\
{\it Albert-Ludwigs-Universit{\"a}t Freiburg,
D-79104 Freiburg, Germany }
\\
${}^b${\it Institut f\"ur Theoretische Teilchenphysik,} \\
  {\it Universit\"at Karlsruhe, D-76128 Karlsruhe, Germany}
 }
\date{}
\maketitle

\begin{abstract} \noindent 
We present analytical results for the leading contributions of the top
quark to the electroweak $\rho$ parameter at order $G_F^3 M_t^6$ and
$G_F^2 M_t^4 \alpha_s$. The Higgs boson and the gauge bosons are taken
to be massless in this limit.  The correction of order $G_F^3 M_t^6$
is found to be sizeable in comparison to the the leading two-loop
${\cal{O}}(G_F^2 M_t^4)$ correction, however it is much smaller than
the subleading ${\cal{O}}(G_F^2 M_t^2 M_Z^2)$ correction. 
\end{abstract}

\thispagestyle{empty}
\setcounter{page}{1}

\renewcommand{\thefootnote}{\arabic{footnote}}
\setcounter{footnote}{0}

\section{Introduction}

The standard model of elementary particle physics has been in agreement with
all experiments performed till date. The one  ingredient that has not been 
found yet is the Higgs boson. Direct searches at LEP200 have been able to put
a lower limit of  $113$  GeV  on its mass \cite{proc1}.  As the presence of the
Higgs boson is necessary for the renormalizability of the theory, radiative 
processes calculated without taking into account the graphs containing virtual 
Higgs bosons would be infinite. Therefore radiative corrections should grow 
with the Higgs-boson mass. It is a well-known effect, sometimes called 
Veltman's theorem \cite{veltact}, that such corrections to low-energy 
parameters grow logarithmically with the Higgs mass. Therefore precision tests
as performed at LEP can use the existing data to put bounds on the Higgs mass.
An analysis of the data indicates that the Higgs boson should be light 
($ <  170$  GeV) \cite{proc2}. Therefore the Higgs boson should be found at 
the next generation of colliders. However there is some lingering doubt as to 
the unavoidability of this conclusion. On the one hand there is a slight 
discrepancy between the leptonic and the hadronic data in the precision 
measurements. Taking the leptonic data only, the preferred value of the Higgs
boson mass would be  so light as to almost rule out the standard model already.
On the other hand one should not  change the model too much, as one can get 
rather large effects quickly, for instance a fourth generation appears to be 
definitely ruled out. The one option that is not quite under theoretical control
is the case of  strong interactions, which corresponds to a naively very heavy
Higgs boson ($M_H \geq 1$ TeV). In this case higher order corrections could be
large and cancel the leading order $\ln(M_H)$  effects, ultimately giving 
similar corrections as a light Higgs boson. Calculations at the two-loop order
have been performed and actually give a negligeable effect \cite{bij1}. The 
reason for this has been traced back to some apparently accidental cancellations
between the coefficients in the Higgs mass dependent terms \cite{haibin}. These
cancellations appear to be typical for the low-energy vector boson-propagator
and triple-vector boson interactions only \cite{bij1}. Corrections to e.g.~the
four-vector boson interactions \cite{jikiaquart} or the Higgs-propagator 
\cite{ghinprop} do not exhibit such cancellations. Therefore it may be possible
that the full effect of the strong interactions shows up only at the three-loop
level for the precision parameters at LEP. It is therefore imperative to push
the calculations towards the three-loop order.

Of course, to determine the Higgs mass effects it is necessary to also have the
effects of the other particles under very precise control, as these may be 
dominating. A particularly important parameter here is the so-called $\rho$ 
parameter which measures the relative strengths of the neutral and charged 
currents. At the tree-level this parameter is one, due to a residual SU(2), 
so-called custodial symmetry. This symmetry gets violated by the hyper-charge 
and by mass-splittings within doublets. Because the mass-splittings give 
contributions proportional to the square of the masses, the dominating effect
is the contribution of the top quark ($\Delta \rho \approx  G_F M_t^2$) 
\cite{rhoone}. The Higgs mass dependence is only logarithmic and proportional to 
the hypercharge coupling ($\Delta \rho \approx g_Y^2 \ln(M_H/M_W)$) 
\cite{longhi}. Since the Higgs mass dependent effect is sub-dominant, it could
be masked by higher loop effects coming from the top quark. Therefore the 
higher loop top mass dependent corrections have to be calculated. As the
top-Yukawa coupling is not very small, one should check whether perturbation 
theory in the Yukawa coupling is actually convergent. Known corrections at the
two-loop level are the leading ($G_F^2 M_t^4$) and sub-leading 
($g_{weak}^2 G_F M_t^2 $) contributions. Numerically it was found that the 
subleading corrections are larger than the leading ones \cite{Degrassi:1996mg}.
This naturally brings up the question whether the three-loop effect could also
be large. There is actually an extra enhancement of the three-loop correction,
because there is a term behaving like $n_c^2$ ($n_c$ is the number of colors),
whereas both one and two-loop corrections behave like $n_c$. If indeed the 
three-loop correction would be comparable to the two-loop one, one should 
consider whether there is a way to resum at least the leading order terms, 
maybe by a Borel summation. Such a resummation is not always possible, as one 
would need an alternating sign in the series. So independent of its direct 
phenomenological implications it is important to know the three-loop heavy-top 
correction to $\Delta \rho$, since it would give us a handle on the nature of 
perturbation theory in quantum field theory. Not very many quantities are known
to such precision.

As far as the calculation is concerned one must remark that a complete
calculation at the three-loop level is impossible analytically. At best this
could be done numerically. However the situation  is somewhat simplified
by taking the top mass much larger than $M_W$ or $M_Z$. In this case one
can perform an expansion in the external momenta. The Feynman integrals that 
have to be calculated can then be reduced to vacuum graphs. These vacuum graphs
are easier to calculate. At the two-loop level this program can be carried out
till the end, as the vacuum-diagrams are known for arbitrary values of the
masses. This way the leading and sub-leading corrections due to a heavy top
quark have been calculated, including a full Higgs mass dependence
\cite{Barbieri92-93,Fleischer95}.  Numerically
the Higgs mass dependence is quite sizeable. Unfortunately at the three-loop
level vacuum diagrams are only known when there is one mass scale in the
problem \cite{broadhurst}. That is, every line in the graph is either massless
or has a fixed mass.  Mixed Higgs-top graphs can therefore only be calculated
when the Higgs has the same mass as the top quark, or has mass zero.  Also 
QCD corrections can be taken into account analytically. In this letter we 
consider the case that the top mass is much heavier than all other masses,
effectively working in the  $M_H=0$  limit.  Within this limit we present the
results of the calculation of the leading order $M_t^6$ and $\alpha_s M_t^4$ 
correction to $\Delta \rho$.

This letter is organized as follows: In the next Section we state our
definition of the $\rho$ parameter and introduce our notations. Section 
\ref{sec:diaren} is devoted to a short discussion of the treatment of the
diagrams and the renormalization. In Section \ref{sec:results} we present 
the results at order $G_F^3 M_t^6$ and $G_F^2 M_t^4 \alpha_s$.

\section{\label{sec:definition}Definition}

The $\rho$ parameter is usually defined by the ratio of the neutral and charged
current coupling constants at zero momentum transfer:
\begin{equation}
\rho = \frac{J_{NC}(0)}{J_{CC}(0)} = \frac{1}{1 - \Delta \rho}.
\label{rho}
\end{equation}
$J_{CC}(0)$ is given by the Fermi coupling constant $G_F$ determined from the
$\mu$ decay rate whereas $J_{NC}(0)$ is measured by neutrino scattering on 
electrons or hadrons. This definition of the $\rho$ parameter is principally 
speaking not process independent, since the radiative corrections depend on 
the value of the hypercharge of the particles in the process. However the 
leading terms in the top mass are process independent.

The origin of the process dependence lies in the different hypercharge
assignments of the particles that take part in the scattering cross-sections.
As already mentioned in the introduction, the origin of a non-zero value of 
$\Delta \rho$ is due to the breaking of the custodial SU(2) symmetry. This 
breaking originates from either the hypercharge coupling or from the doublet 
mass-splittings. In leading order these effects are separate, but in higher 
order the effects of the Yukawa and the gauge couplings become intricately 
intertwined. At the two-loop level there is a leading ${\cal O}(M_t^4)$ 
correction, that is independent of the gauge-couplings completely. Even in the 
absence of the hypercharge field there is a subleading ${\cal O}(M_t^2)$ 
contribution. Finally the exact sub-leading term becomes a function of the 
hypercharge coupling, giving rise to a more complicated formula as a function
of $\sin^2(\theta_W)$ \cite{Degrassi:1996mg}. As the whole correction is also 
strongly dependent on the Higgs mass one should be careful in drawing 
conclusions from the large size of the subleading term compared to the leading
term. This effect may be specific to the particular observable being 
calculated. In particular there appears to be an accidental cancellation 
between different coefficients in the leading term ${\cal O}(G_F^2 M_t^4)$ to 
$\Delta \rho$ for $M_H=0$, which disappears for a larger Higgs mass and which
is absent in the correction to $Z \rightarrow b\bar b$.

In leading order in the top mass contributions to $\rho$ stem from the
transversal parts of the (unrenormalized) self energies of the exchanged vector 
bosons $W$ and $Z$:
\begin{equation}
\rho = \frac{1 - \Pi_T^{WW}(0)/M_{W_0}^2}{1 - \Pi_T^{ZZ}(0)/M_{Z_0}^2}
{},
\end{equation}
where $M_{W_0}\, M_{Z_0}$ are the bare masses of the $W$ and $Z$ bosons 
respectively. In this form the results would appear infinite, but these 
infinities are canceled by the Higgs mass, top mass and $W$ boson mass 
renormalizations. Corrections from vertex and box diagrams always involve extra
powers of the weak coupling constant $g_{weak}$ and are therefore suppressed by
powers of $g_{weak}/g_{Yukawa}^{top}$, i.e.~$M_W/M_t$.

A remark is here in order. Up to the two-loop level for the leading top mass 
effects the correction to $\rho$ can be expressed in a number of equivalent 
ways, i.e.~in terms of $G_F^2 M_t^4$ or $M_t^4/M_W^4$. At the three-loop level 
one must be more precise. At this level the difference between low-energy 
parameters and on-shell vector boson masses plays a role even in the leading 
$M_t$ effects. In the following we will express the final answer in terms of 
the on-shell top mass and the low-energy Fermi constant $G_F$ from $\mu$-decay.

\section{\label{sec:diaren}Treatment of diagrams and renormalization}
 
There are four major contributions to the three-loop $\rho$ parameter which 
we will discuss below. First, the three-loop self energy diagrams of the $W$ 
and $Z$ boson, of course. Second, the top quark mass renormalization at order 
$G_F^2 M_t^4$ and $G_F M_t^2 \alpha_s$. Third, the renormalization of the 
vacuum expectation value of the Higgs field and fourth, the Higgs mass 
renormalization. For all contributions we have limited ourselves to the leading
top quark mass effects which simplifies the computation drastically.

As far as the self energy diagrams of the $W$ and $Z$ bosons are concerned we 
have chosen to compute an asymptotic expansion w.r.t.~the external momentum 
since we are interested in the zero momentum transfer limit. The treatment of
the Higgs mass has been done using two different approaches: First, an 
asymptotic expansion \cite{Smi95} w.r.t.~the Higgs mass and second, setting 
the Higgs mass equal to zero right from the start. Both strategies lead to 
the same result as no Higgs mass dependence is present in the final results. 

Throughout this computation we have used several computer programs and
packages. The diagrams were generated using {\tt QGRAF} \cite{QGRAF}
and the asymptotic expansion in limit of the small Higgs mass was performed 
by {\tt EXP} \cite{EXP}. The resulting three-loop integrals were evaluated 
with the help of the program packages {\tt MINCER} \cite{MINCER} and 
{\tt MATAD} \cite{MATAD} written in {\tt FORM} \cite{FORM}. The two-loop top
quark mass renormalization constant requires the calculation of two-loop 
on-shell diagrams. We have used the packages {\tt TARCER} \cite{TARCER} and
{\tt ONSHELL2} \cite{ONSHELL2} for the computation of these types of diagrams. 

The renormalization of the vacuum expectation value of the Higgs field is fixed
by the requirement of vanishing  Higgs tadpoles. It also influences the Higgs
mass renormalization and the mass renormalization of the Goldstone bosons. We 
need this renormalization constant up to two-loops and the computation of the
tadpoles is straightforward. The mass renormalization of the Higgs field is 
required only up to one-loop order.

\begin{figure}[t]
  \begin{center}
    \begin{tabular}{cc}
      \leavevmode
      \epsfxsize=5.cm
      \epsffile{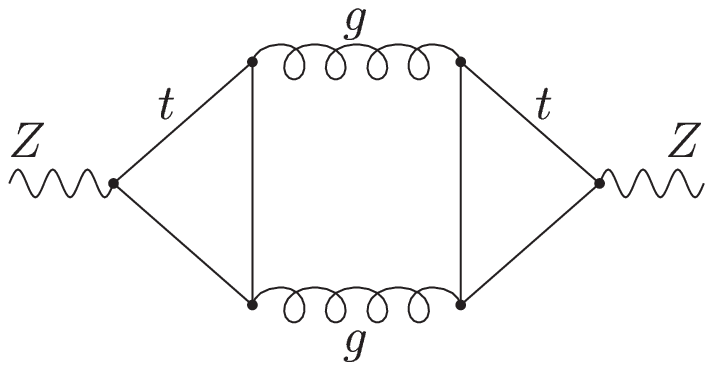} &
\leavevmode
      \epsfxsize=5.cm
      \epsffile{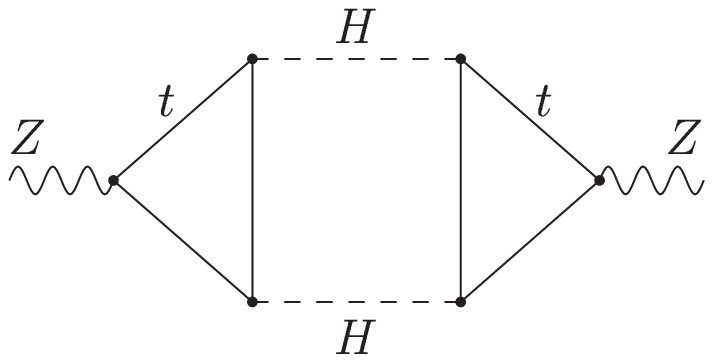}
\\
(a)  & (b)
    \end{tabular}
  \end{center}
  \caption{\label{fig:diags}(a) A diagram for which the traces with an odd
number   of $\gamma_5$ matrixes {\em do} contribute; (b) a diagram which can be
safely    treated with  the completely anticommutating $\gamma_5$.}
\end{figure}

We have used the prescription of a completely anticommutating $\gamma_5$ in 
the evaluation of all traces. In particular, all traces  with an odd number
of $\gamma_5$'s were set to zero. This treatment is completely justified if
and only if every such diagram vanishes on kinematical grounds. For instance,
obviously no problem could result from 2-point amplitudes with only one fermion
loop: the number of available independent external momenta (one) and Lorentz 
indices (two if both legs are vector bosons) is not enough to form a completely
ntisymmetric tensor $\epsilon_{\mu \nu \lambda \rho }$. The situation can be
different only for singlet type 3-loop diagrams comprising two fermion traces 
(see Fig. 1).

Indeed, the trace occurring in one-loop vertex diagram with three vector 
external legs (Fig.~1(a)) can only be computed by using the 't~Hooft-Veltman 
prescription for $\gamma_5$. Fortunately, in our case the only dangerous 
diagrams are those of Fig.~1(b). They contain at worst triangle subdiagrams 
comprising a trace of an odd number of $\gamma_5$'s and depending on only 
three (not four!) independent external Lorentz vectors and indices (two 
independent external momenta and at maximum one index from a vector boson leg).

We have also successfully checked the Ward-Takahashi identities which relate 
the self-energies of the gauge bosons to the non-diagonal self-energies of the
gauge-to-Goldstone and the Goldstone boson self-energies.

\section{\label{sec:results}Results}

We have reproduced the one- and two-loop results for the $\rho$ parameter at
orders $G_F M_t^2$, $G_F^2 M_t^4$ and $G_F M_t^2 \alpha_s$ that can be found in
\cite{rhoone,rhooneals,rhotwo}:
\begin{eqnarray}
\Delta \rho^{(1,M_t^2)} &=& n_c X_t \label{drho1}
{}\, ,
\\
\Delta \rho^{(2,M_t^4)} &=& n_c X_t^2 \left( 19 - 12 \zeta_2 \right)
\label{drho2}
{}\, ,
\\
\Delta \rho^{(2,\alpha_sM_t^2)} &=&  n_c X_t \frac{\alpha_s}{\pi} C_F 
                      \left( - \frac{1}{2} - \zeta_2 \right)
{}\, .
\label{drho2alphas}
\end{eqnarray}
$X_t$ is defined by ($M_t$ is the top quark pole mass)
\begin{equation}
X_t = \frac{G_F M_t^2}{8 \sqrt{2} \pi^2}
\end{equation}
and $n_c$ represents the number of colours. $\zeta_n$ is used for the values of
the Riemann $\zeta$-function at $n$, $\zeta(n)$, giving 
e.g.~$\zeta_2 = \pi^2/6$. $C_F$ is equal to $4/3$ for the SU(3) colour group.

For the light Higgs boson the subleading two-loop ${\cal O}(G_F^2M_t^2M_Z^2)$
corrections are much larger than the leading two-loop ${\cal O}(G_F^2M_t^4)$
term, which is actually suppressed by accidental cancellations 
\cite{Degrassi:1996mg}. So they should necessarily be taken into account. In
contrast to the leading corrections the subleading $M_t^2$ two-loop corrections
depend on the processes used to define the neutral and charged currents in 
Eq.~(\ref{rho}). If the charged current coupling constant is defined through
the $\mu$-decay and the neutral current coupling constant is defined through
the $\nu_\mu e$ scattering, then the subleading two-loop corrections in the 
massless Higgs limit are given by
\begin{eqnarray}
\Delta\rho^{(2,M_t^2)}&=&
n_c X_t^2 \, \frac{M_Z^2}{M_t^2} \,
\Biggl[ - \frac{11}{2} + \frac{3}{\hat{s}^2} + \frac{319\,\hat{s}^2}{9} 
 +  6\,\hat{c}^2\,I_3  + {\pi }^2\,\left( -\frac{7}{3} 
- \frac{56\,\hat{s}^2}{27}\right)\nonumber \\ 
&&+ \left( 7 + \frac{3}{\hat{s}^4} - \frac{6}{\hat{s}^2} - 4\,\hat{s}^2 \right)
       \,\ln (\hat{c}^2) + \left( 21 - 16\,\hat{s}^2 \right) 
\,\ln \left( \frac{M_Z^2}{M_t^2} \right) \Biggr],
\label{drho21}
\end{eqnarray}
with $I_3$ representing the isospin of the electron target ($I_3=-1$) 
\cite{Degrassi:1996mg}. $\hat{s}$ and $\hat{c}$ are abbreviations for the 
$\overline{\mbox{MS}}$ quantities $\sin\hat{\theta}_W$ and $\cos\hat{\theta}_W$,
respectively, given by $\hat{c}^2\approx c^2(1-n_c X_t)$ with the on-shell 
parameter $c^2=M_W^2/M_Z^2$. In the limit of vanishing hypercharge ($\hat{s}=0$)
the corresponding correction looks like 
\begin{equation}
\Delta\rho^{(2,M_t^2)}\Bigg|_{\hat{s}=0}=
 n_c X_t^2 \frac{M_Z^2}{M_t^2} \,\Biggl[ -1 - \frac{7\,{\pi }^2}{3} 
 + 6\,I_3 + 21\,\ln \left( \frac{M_Z^2}{M_t^2} \right) \Biggr].
\label{drho210}
\end{equation}
Numerically the subleading corrections (\ref{drho21}-\ref{drho210}) are about
20 times larger than the leading two-loop correction (\ref{drho2}) and are 
about 4\% of the one-loop correction (\ref{drho1}).

\begin{table}[t]
\begin{center}
\begin{tabular}{|c|c|c|c|c|c|c|}
\hline
$X_t$ & $X^2_t$ & $X_t^2\, \frac{M_Z^2}{M_t^2}$ & $X_t\alpha_s$ & $X^3_t$ &  $X^2_t\alpha_s$ & 
$X_t\alpha^2_s$  
\\
\hline
$ 9.6 \cdot 10^{-3}$ & $ -2.3 \cdot 10^{-5} $ & $-4.1 \cdot 10^{-4}$  & $-9.5 \cdot 10^{-4}$   & $8.2 \cdot 10^{-6}$  
& 
$1.0 \cdot 10^{-6}$   &  $-1.7 \cdot 10^{-4}$      
\\
\hline
\end{tabular}
\caption{Size of different contributions to the $\rho$ parameter according to 
Eq.~(\protect\ref{result}). The numbers 
are obtained with $\alpha_s(M_t) = 0.109$.}
\end{center}
\end{table}

The  three-loop results read
\begin{eqnarray}
\lefteqn{\Delta \rho^{(3,M_t^6)} =} \nonumber\\
&&\mbox{} X_t^3 n_c 
  \left( 68 + 729 S_2 + 36 D_3 + 96 \zeta_2 \ln 2 + 6 \zeta_2 
         - 612 \zeta_3 + 324 \zeta_4 - 72 B_4 \right) \nonumber\\&&\mbox{}
  + X_t^3 n_c^2 \left( - \frac{6572}{15} - \frac{4374}{5} S_2 
                       + \frac{1472}{15} \zeta_2 + 440 \zeta_3 \right)
\label{drho3}
\end{eqnarray}
and
\begin{eqnarray}
\lefteqn{\Delta \rho^{(3,M_t^4 \alpha_s)} =} \nonumber\\
&&\mbox{}
X_t^2 n_c \frac{\alpha_s}{\pi} C_F
\left(
  \frac{185}{3} + \frac{729}{4} S_2 - 48 \zeta_2 \ln 2 
  - \frac{151}{6} \zeta_2 + 29 \zeta_3 - 24 \zeta_4 + 12 B_4
  \right).
\label{drho3alphas}
\end{eqnarray}
The constants $S_2$, $D_3$, and $B_4$ are given by (see, e.g. \cite{MATAD})
\begin{eqnarray}
  S_2 &=& \frac{4}{ 9 \sqrt{3} } \mbox{Cl}_2 \left( \frac{\pi}{3} \right)
  = 0.260434\dots
  \nonumber\\
  D_3 &=&
  6 \zeta_3
  - \frac{15}{4}\zeta_4 
  - 6 \left[ \mbox{Cl}_2 \left( \frac{\pi}{3} \right) \right]^2 
  = -3.027009\dots
  \nonumber\\
  B_4 &=& 
  16 \mbox{Li}_4 \left( \frac{1}{2} \right) 
  - 4 \zeta_2 \ln^2 2
  + \frac{2}{3} \ln^4 2 
  - \frac{13}{2} \zeta_4
  = -1.762800\dots
\end{eqnarray}
with $\mbox{Cl}_2(\pi/3) = \mbox{Im}[\mbox{Li}_2(e^{i\pi/3})]$. $\mbox{Li}_2$ and 
$\mbox{Li}_4$ represent the Di- and Quadrilogarithm, respectively.

Numerically, we find 
\begin{eqnarray}
\Delta \rho &=& n_c \, X_t 
                - 0.7392 n_c \, X_t^2 
                -X_t^2 \, \frac{M_Z^2}{M_t^2} \, n_c \,
                \left(30.029 -21 \ln \left( \frac{M_Z^2}{M_t^2} \right)
                 \right)
\nonumber \\ 
&&\mbox{}
                + n_c ( 10.1466  + 24.3669 \, n_c) \, X_t^3\nonumber \\
&&\mbox{}
                + \frac{\alpha_s}{\pi} n_c \left( 
                - 2.8599 \, X_t 
                + 0.9798 \, X_t^2 
                \right)  
                + \cdots
\label{result:nc}
\end{eqnarray}
and, for $n_c=3$, 
\begin{eqnarray}
\Delta \rho &=& 3 \, X_t 
                - 2.2176 \, X_t^2 
                -X_t^2 \, \frac{M_Z^2}{M_t^2} 
                 \left( 90.1 - 63 \ln \left( \frac{M_Z^2}{M_t^2} \right) 
                 \right)
                + 249.74 \, X_t^3\nonumber \\
&&\mbox{}
                + \frac{\alpha_s}{\pi} \left( 
                 -8.5797 \, X_t 
                 + 2.9394 \, X_t^2 
                \right)                
\nonumber \\ 
&&\mbox{}
- 43.782 \left(\frac{\alpha_s}{\pi}\right)^2 X_t  +  \cdots
\label{result}
\end{eqnarray}
Here the subleading two-loop correction is given again for the process of 
$\nu_\mu e$ scattering and in order to see the pure effect of the top quark 
mass on $\Delta\rho$ the value $\hat{s}=0$ was taken. For the realistic value 
of $\hat{s}^2=0.2311$ the subleading two-loop corrections are 
$-X_t^2 M_Z^2/M_t^2 \, ( 78.7-51.9\ln(M_Z^2/M_t^2) )$. For completeness we have
also added to the r.h.s.~of (\ref{result}) the well known term of order 
$G_F M_t^2 \alpha_s^2$ \cite{tarasov}.

The size of the various terms in Eq.~(\ref{result}) is shown in Table 1. A few
remarks are in order. First, the coefficients of the perturbative series are 
alternating for both pure electroweak and mixed electroweak/strong corrections. 
Second, the size of the three-loop electroweak coefficient is quite large and 
results in a correction to the leading two-loop contribution of approximately
36 \% (for $X_t \approx 3.2 \cdot 10^{-3}$). The size of the three-loop 
correction relative to the much larger subleading two-loop $M_t^2$ corrections
(\ref{drho21}-\ref{drho210}) is however about 2\%. The mixed 
${\cal{O}}(G_F^2 M_t^4 \alpha_s)$ term is numerically less than the purely 
electroweak ${\cal{O}}(G_F^3 M_t^6)$ one by almost an order of magnitude. 
Finally, the major part of the ${\cal O}(X_t^3)$ contribution to the $\rho$ 
parameter is given by the leading term in the large $n_c$ approximation.

With respect to the question of the convergence of perturbation theory in the
Yukawa-coupling $X_t$ the results give only limited information. Because of the
dependence on $n_c$ one can actually study two different limits, selecting 
different subsets of graphs. The first possibility is to take the graphs with
the largest number of fermion-loops into account, which is one less than the 
number of loops, starting at the two-loop level. Formally one can project out
these graphs by taking $X_t \approx 1/n_c$ and take the limit $n_c \to \infty$.
With the exception of one-loop all other loops are of the same order. One can 
thus only meaningfully compare the two-loop with the three-loop level. These 
become equal for $n_c X_t \approx 0.03$, corresponding to $M_t \approx 300  \, $
GeV. This would naively imply an early breakdown of perturbation theory. Such a
conclusion however would be premature, as it is known that the two-loop 
$\rho$ parameter has a large cancellation between the different terms, that 
have not been observed for other quantities. 

The other limit is taking into account only graphs with one fermion-loop,
which one can describe formally by taking the limit $n_c \rightarrow 0$.
Here one has three terms in the perturbation expansion available and can 
therefore form the [1,1]-Pad\'e  approximant.  This gives
\begin{equation}
\Delta \rho = n_c X_t \frac{1 + 12.987 X_t}{1 + 13.7265 X_t}
\, .
\end{equation}
For a large top mass ($X_t \to \infty$) this would imply a saturation of the 
radiative correction 
\begin{equation}
 \Delta \rho = 0.946 \,\Delta \rho_{1-loop}\,.
\end{equation}
However also this early saturation could be an artefact, coming from the 
anomalously small two-loop leading term. All in all one gains little insight
here without having a way to sum high-loop graphs. Also a comparison with other 
quantities would be necessary as well as a study of the Higgs mass dependence.

\section*{Conclusion} 

We have analytically computed the leading contributions of the top quark to the
electroweak $\rho$ parameter at order $G_F^3M_t^6$ and $G_F^2M_t^2\alpha_s$. 
The correction of order $G_F^3M_t^6$ is sizeable in comparison to the leading
two-loop correction. As we have mentioned in the Introduction the three-loop 
leading correction in principle could have happened to be of the same order of
magnitude as the subleading $G_F^2M_t^2M_Z^2$ two-loop correction. We have 
found, however, that in fact it is still much smaller than the subleading 
two-loop one. Thereby we have significantly reduced the theoretical uncertainty
of the results of electroweak higher order calculations coming from the 
enhanced three-loop large top quark mass contributions.
 
\section*{Acknowledgements} 

This work was supported by the {\it Graduiertenkolleg
``Elementarteilchenphysik an Beschleunigern''} and the {\it
DFG-Forschergruppe ``Quantenfeldtheorie, Computeralgebra und
Monte-Carlo-Simulation''} ( contract FOR 264/2-1). It was also
supported by the European Union under contract HPRN-CT-2000-00149.

The authors would like to thank J.H.~K\"uhn for useful discussions
and advice.  We are grateful to M.Yu.~Kalmykov for providing us with
an updated version of the package {\tt ONSHELL2}.

\def\app#1#2#3{{\it Act.~Phys.~Pol.~}{\bf B #1} (#2) #3}
\def\apa#1#2#3{{\it Act.~Phys.~Austr.~}{\bf#1} (#2) #3}
\def\cmp#1#2#3{{\it Comm.~Math.~Phys.~}{\bf #1} (#2) #3}
\def\cpc#1#2#3{{\it Comp.~Phys.~Commun.~}{\bf #1} (#2) #3}
\def\epjc#1#2#3{{\it Eur.\ Phys.\ J.\ }{\bf C #1} (#2) #3}
\def\fortp#1#2#3{{\it Fortschr.~Phys.~}{\bf#1} (#2) #3}
\def\ijmpc#1#2#3{{\it Int.~J.~Mod.~Phys.~}{\bf C #1} (#2) #3}
\def\ijmpa#1#2#3{{\it Int.~J.~Mod.~Phys.~}{\bf A #1} (#2) #3}
\def\jcp#1#2#3{{\it J.~Comp.~Phys.~}{\bf #1} (#2) #3}
\def\jetp#1#2#3{{\it JETP~Lett.~}{\bf #1} (#2) #3}
\def\mpl#1#2#3{{\it Mod.~Phys.~Lett.~}{\bf A #1} (#2) #3}
\def\nima#1#2#3{{\it Nucl.~Inst.~Meth.~}{\bf A #1} (#2) #3}
\def\npb#1#2#3{{\it Nucl.~Phys.~}{\bf B #1} (#2) #3}
\def\nca#1#2#3{{\it Nuovo~Cim.~}{\bf #1A} (#2) #3}
\def\plb#1#2#3{{\it Phys.~Lett.~}{\bf B #1} (#2) #3}
\def\prc#1#2#3{{\it Phys.~Reports }{\bf #1} (#2) #3}
\def\prd#1#2#3{{\it Phys.~Rev.~}{\bf D #1} (#2) #3}
\def\pR#1#2#3{{\it Phys.~Rev.~}{\bf #1} (#2) #3}
\def\prl#1#2#3{{\it Phys.~Rev.~Lett.~}{\bf #1} (#2) #3}
\def\pr#1#2#3{{\it Phys.~Reports }{\bf #1} (#2) #3}
\def\ptp#1#2#3{{\it Prog.~Theor.~Phys.~}{\bf #1} (#2) #3}
\def\sovnp#1#2#3{{\it Sov.~J.~Nucl.~Phys.~}{\bf #1} (#2) #3}
\def\tmf#1#2#3{{\it Teor.~Mat.~Fiz.~}{\bf #1} (#2) #3}
\def\yadfiz#1#2#3{{\it Yad.~Fiz.~}{\bf #1} (#2) #3}
\def\zpc#1#2#3{{\it Z.~Phys.~}{\bf C #1} (#2) #3}
\def\ppnp#1#2#3{{\it Prog.~Part.~Nucl.~Phys.~}{\bf #1} (#2) #3}
\def\ibid#1#2#3{{ibid.~}{\bf #1} (#2) #3}
\def\jhep#1#2#3{{\it JHEP~}{\bf #1} (#2) #3}

\end{document}